# Accelerated Carrier Relaxation through Reduced Coulomb Screening in 2D Halide Perovskite Nanoplatelets

Verena A. Hintermayr[1,2], Lakshminarayana Polavarapu[1,2], Alexander S. Urban[1,2,3*], Jochen Feldmann[1,2*]

[1] *Chair for Photonics and Optoelectronics, Department of Physics, Ludwig-Maximilians-Universität München, Amalienstr. 54, 80799 Munich, Germany.*

[2] *Nanosystems Initiative Munich (NIM) and Center for NanoScience (CeNS), Schellingstr. 4, 80799 Munich, Germany.*

[3] current address: *Nanospectroscopy Group, Department of Physics, Ludwig-Maximilians-Universität München, Amalienstr. 54, 80799 Munich, Germany*

e-mail: urban@lmu.de; feldmann@lmu.de

## Abstract

For high-speed optoelectronic applications relying on fast relaxation or energy transfer mechanisms, understanding of carrier relaxation and recombination dynamics is critical. Here, we compare the differences in photoexcited carrier dynamics in 2D and quasi-3D colloidal methylammonium lead iodide perovskite nanoplatelets *via* differential transmission spectroscopy. We find that the cooling of excited electron-hole pairs by phonon emission progresses much faster and is intensity-independent in the 2D-case. This is due to the low dielectric surrounding of the thin perovskite layers, for which the Fröhlich interaction is screened less efficiently leading to higher and less density dependent carrier-phonon scattering rates. In addition, rapid dissipation of heat into the surrounding occurs due to the high surface-to-volume ratio. Furthermore, we observe a sub-picosecond dissociation of resonantly excited 1s excitons in the quasi-3D case, an effect which is suppressed in the 2D nanoplatelets due to their large exciton binding energies. The results highlight the importance of the surrounding environment of the inorganic nanoplatelets on their relaxation dynamics. Moreover, this 2D material with relaxation times in the sub-picosecond regime shows great potential for realizing devices such as photodetectors or all-optical switches operating at THz frequencies.

**Keywords: perovskite, nanoplatelets, coulomb screening, carrier relaxation, transient absorption spectroscopy, Fröhlich interaction,**

Lead halide perovskite thin films have established themselves as an excellent material system for photovoltaic applications due to favorable absorption and charge transport properties.[1-4] With bandgaps tunable throughout the visible range *via* halide-ion replacement [5,6] and quantum yields approaching unity, perovskite nanocrystals (NCs) exhibit strong potential for a variety of other optoelectronic applications.[7-10] While in both of these fields either bulk or only weakly-confined perovskite materials have been employed, recently strongly-confined perovskite NCs, such as 2D nanoplatelets (NPls) with a monolayer-precise control of the thickness and resulting quantum size effects have emerged.[11-15] With extremely high exciton binding energies and transition energies tunable through quantum-confinement, they demonstrate intriguing possibilities for light-emitting applications such as LEDs or lasers.[16-18] Moreover, with excitons dominant at room-temperature perovskite NPls are also promising for excitonic device concepts. Importantly, high-frequency limitations depend on the timescales for relaxation and recombination scenarios of photoexcited or electrically injected charge carriers.

In this paper, we investigate charge carrier relaxation in 2D and quasi-3D methylammonium lead iodide (MAPI) NPls by means of femtosecond differential transmission spectroscopy (DTS) at room temperature. Linear absorption spectroscopy initially reveals the effect of quantum-confinement on the continuum absorption onset position and on the exciton binding energy, which is found to be more than ten times larger for the 2D NPls as compared to the quasi-3D NPls. Consistently, results of transient DTS measurements show a bimolecular recombination due to free electron-hole pairs for the quasi-3D NPls and a monomolecular decay behavior characteristic for excitonic decay for the 2D NPls. Regarding the initial carrier relaxation we find an intensity-independent sub-picosecond cooling time in the 2D NPls which is seven times faster than the strongly intensity-dependent cooling time in the quasi-3D NPls. This can be explained by the large surface-to-volume ratio of the 2D NPls hindering the creation of hot phonons and a less efficient screening of the photoexcited carriers, leading to high exciton-optical phonon scattering rates, governed by the Fröhlich interaction.[19,20] Finally, by comparing resonant and non-resonant pump pulse excitation of the 1s exciton transitions, we directly monitor the exciton dissociation for the quasi-3D case, whereas resonantly excited excitons in the 2D NPls stay on the 1s exciton parabola. Our findings highlight the vast differences occurring in perovskite materials based on their dimensionality. Knowledge of these processes could prove critical to designing optoelectronic devices relying on fast relaxation, charge or energy transfer mechanisms.

## Results

**Confinement effects and exciton binding energies**

MAPI NPls were synthesized by ligand-assisted exfoliation, as previously developed by our group and as detailed in the *Methods* section.[12] Their thickness is quantized according to an integer number of 2D layers of corner-sharing [PbI$_6$]$^{-4}$ octahedra down to a single monolayer (0.66 nm). Resulting NPI dispersions can be used to study the effect of quantum-confinement in one dimension, as the thick bulk-like NPls are progressively shrunk down to the limit of 2D NPls. To show the difference between 2D and 3D NPls, we selected a dispersion of trilayer MAPI NPls (thickness $d \approx 2\ nm$) and thicker NPls ($d \geq 15\ nm$), which are referred to as quasi-3D NPls, due to their bulk-like optical properties (see also Supporting Information, Figures S1 and S2). The thickness of the MAPI trilayer is in the range of the 3D excitonic Bohr radius of 2.5nm, and thus can be considered as quantum confined in one dimension, whereas the bulk-like NPls are unconfined or only weakly confined, hence justifying the distinction as 2D and quasi-3D NPls, respectively.[21]

All measurements presented here were conducted at room temperature on dispersions of colloidal NPls in toluene. In Figure 1a the absorption and photoluminescence (PL) spectra are shown for the 2D and the quasi-3D NPls. The absorption spectrum in the 2D case reveals a 1s exciton resonance at 2.0eV easily distinguishable from the onset of the electron-hole continuum transitions at 2.27 eV. For the quasi-3D case the excitonic transitions spectrally overlap with the continuum absorption onset due to homogeneous and inhomogeneous broadening effects.[22] Applying the Elliott model to the linear absorption spectra of the NPls (cf. Figure S3) we obtain band gaps of $E_g^{3D} = 1.66\ eV$ and $E_g^{2D} = 2.27\ eV$ as well as exciton binding energies of $E_b^{3D} = 19\ meV$ and $E_b^{2D} = 230\ meV$. These values correspond well to the range of previously published values and confirm the bulk-like nature of the thicker platelets.[23-24] For the 3D NPls the exact determination of the binding energy is difficult due the broadening of the absorption spectrum and so the obtained value constitutes an upper bound.[25] Nevertheless, the exciton binding energy of the 2D NPls is more than 10 times as large as in the 3D case. This is caused by a change in dimensionality (in the ideal case: $E_b^{2D} = 4 \cdot E_b^{3D}$) and by a reduced screening of the Coulomb interaction between electron and hole due to the low dielectric constant of the organic ligands and solvent surrounding the inorganic NPls.[8] Similar values for the exciton binding energy were already observed in the 1990s for layered PbI-based perovskite structures.[26-28] The PL spectra of the NPls display single and narrow resonances at $E_{PL}^{3D}$ = 1.62 eV and $E_{PL}^{2D}$ = 2.03 eV, respectively, exhibiting Stokes shifts of less than 20 meV. This confirms the high quality and uniformity of the NPls.[11]

These findings are represented schematically in Figure 1b, where the dispersion relations E(K) for bound and unbound electron-hole pairs are depicted for the two perovskite NC systems. The huge blueshift of approximately 600 meV of the energetic states due to quantum confinement in z-direction, dispersion in only two directions (x,y) and the large exciton binding energy are the prominent characteristics of the 2D NPls. While for the quasi-3D system, the exciton binding energy (19 meV) is smaller than the thermal energy kT at room temperature (~ 26 meV), the value for the 2D system is far above it. Importantly, for 2D NPls it is not enough just to consider quantum confinement as done for quantum dots, since apart from the confinement in z-direction the dispersion in x- and y-directions is still retained, a fact which is necessary to understand the time-resolved optical experiments presented in this paper.

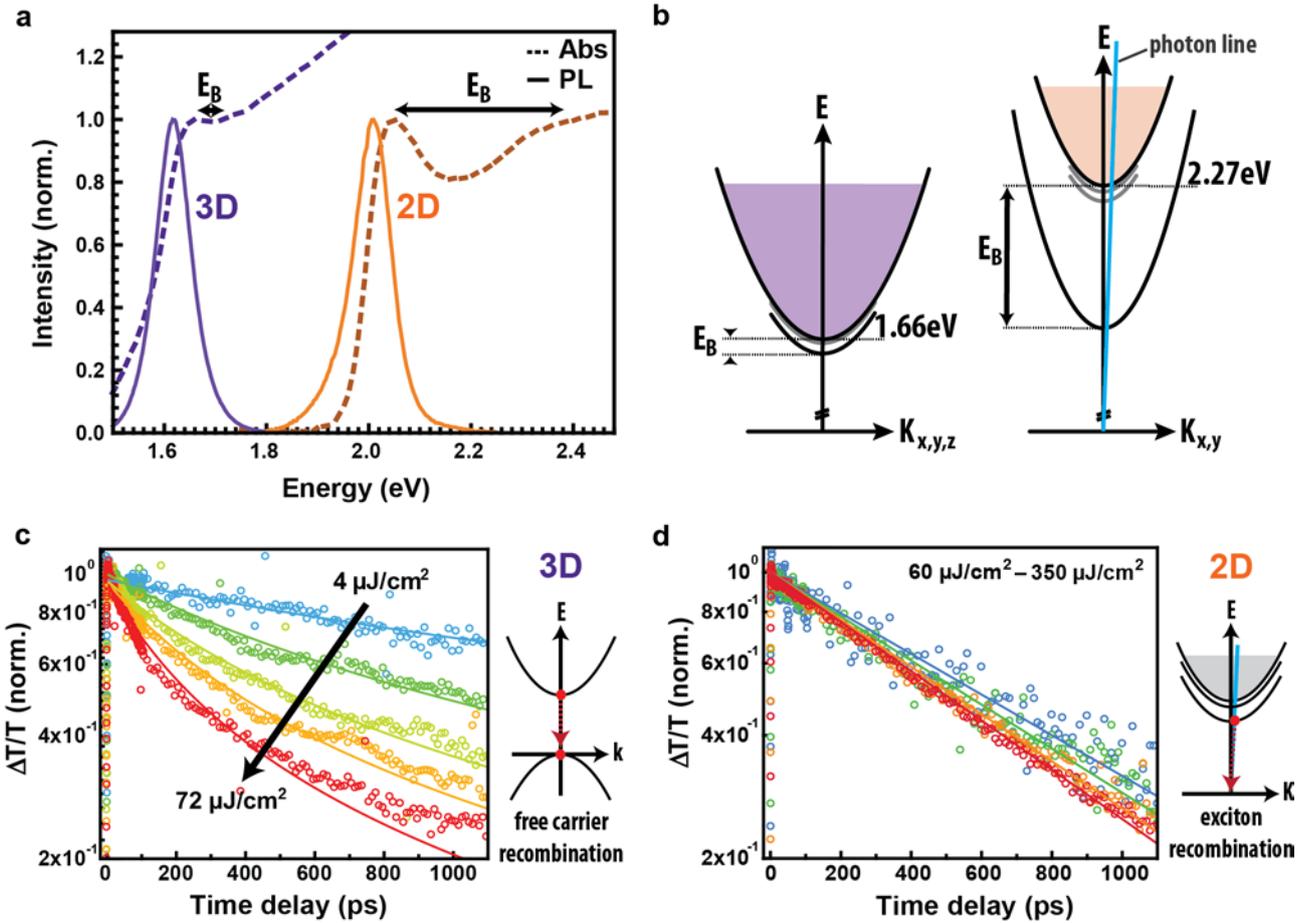

**Figure 1 | Energetics and charge carrier recombination of 2D and quasi-3D MAPI NPls. a,** Steady-state absorption and PL spectra of quasi-3D (purple) and 2D (orange) NPls. The corresponding values of the exciton binding energies $E_B$ are shown as arrows in the absorption spectra. **b,** Dispersion relations $E(K)$ for bound and unbound electron-hole pairs. Within this two-particle picture $K$ represents the wave-vector for the center-of-mass motion of electron-hole pairs. **c,d,** Transients of the ΔT/T signals at the 1s exciton for the quasi-3D **(c)** and 2D case **(d)** as a function of laser excitation density (here shown for a monolayer nanoplatelet). The solid lines are the results of calculated transients assuming bimolecular decay kinetics (ΔT ~ $n_e \times n_h$) and monomolecular decay kinetics (ΔT ~ n) for the 3D and 2D case, respectively. On the right side the corresponding recombination process is depicted for the quasi-3D in the one-particle **(c)** and for the 2D-case in the two-particle **(d)** picture, respectively.

**Photoinduced carrier dynamics**

Time-resolved optical studies on relaxation and recombination scenarios of photoexcited electrons and holes in halide perovskite nanoplatelets have been restricted to PL to date. As indicated by the photon line for the 2D case in Fig.1b in such non-resonant PL experiments electron-hole pairs are typically excited far above the continuum edge and then relax down to the 1s exciton parabola. Radiative emission finally takes place from the crossing point of the 1s exciton parabola $E_{1s}(K_{x,y})$ with the photon line close to K=0. Due to a finite homogeneous linewidth the radiative emission stems from a respective "radiative window" ΔE corresponding to an area ΔK in $K_{x,y}$-space and to a coherence area in real space as theoretically introduced and experimentally demonstrated for the case of GaAs quantum wells [29] and recently also shown for II-VI nanoplatelets.[30]

Consequently, time-resolved PL experiments can provide nearly no information on the initial relaxation dynamics, which occurs before the electron-hole pairs reach the radiative window on the 1s exciton parabola. Instead, transient absorption experiments with femtosecond pump and probe laser pulses have proven to be sensitive to the initial ultrafast dynamics of photoexcited electron-hole pairs.[31] The photoexcitation by the pump pulse initially creates a coherent and highly non-thermal electron-hole distribution at the $E(K)$-position, where the photon line meets the continuum absorption for the given photon excitation energy (Fig. 2a). This non-thermal distribution then thermalizes typically on a sub-picosecond time-scale through carrier-carrier and carrier-phonon scattering.[31-32] Thermalisation means that thereafter the energetic distribution of electrons and holes can be described with Fermi-Dirac distribution functions $f_{e,h}(E,T_c)$. The distribution is generally referred to as "hot", as carrier temperatures $T_c$ can initially be higher than 1000 K. Through phonon emission these carrier distributions cool down until the charge carriers reach thermal equilibrium with the crystal lattice. With sufficient excess energy of the charge carriers the initial cooling occurs within a few picoseconds due to Coulomb-mediated scattering with optical phonons (Fröhlich-interaction). The cooling process typically slows down for higher excitation densities, as phonon reabsorption increases with increasing phonon density ("hot phonon effect")[33,34] and the high density of photoexcited carriers leads to a screening of the Fröhlich-interaction.[19,20,31,35] Once the carrier distribution has cooled so far that optical phonons cannot be emitted anymore ($kT_C \lesssim E_{LO}$), the final cooling can only progress through less efficient acoustic phonon scattering, typically caused by deformation potentials.[20,36] All these relaxation and cooling processes influence the onset and (if slow enough) even the decay behavior of the 1s exciton PL.

**Recombination**

To investigate the relaxation and recombination, we have employed transient DTS on the 2D and 3D NPls, exciting far above the continuum onset at 400 nm, focusing initially on the decay of the ΔT signal for long time delays, τ between the pump and probe pulses. As shown in Figures 1c and 1d, the observed ΔT(τ) transients for 2D and quasi-3D NPls behave differently. For the quasi-3D NPls the decay at the position of maximum bleaching signal follows a $1/\tau$ behavior, indicating bimolecular decay kinetics, and with $\Delta T \sim n_e \times n_h$ likely stemming from free carrier recombination. The decay becomes progressively faster as the excitation density in increased. In contrast, the decay behavior of the $\Delta T(\tau)$ transients at the position of the largest bleaching signal is purely exponential for 2D NPls indicating monomolecular decay kinetics, and with $\Delta T \sim n$, likely excitonic recombination (Fig. 1d). Excitonic PL transients of halide perovskite mono-, bi- and tri-layers also show such a monomolecular decay behavior as recently discussed by our group.[12] This is a direct consequence of the significantly enhanced exciton binding energy in the 2D NPls, which impedes a thermal dissociation of the exciton at room-temperature. In contrast, in the quasi-3D NPls, with an exciton binding energy below the thermal energy at room temperature, free electrons and holes are predominant. In both cases as the excitation density is increased, the initial decay of the measured ΔT transients becomes slightly faster than described by the calculated transients. Auger recombination and exciton-exciton annihilation set in for the 3D and 2D NPls, respectively, as already reported for bulk halide perovskite films.[23]

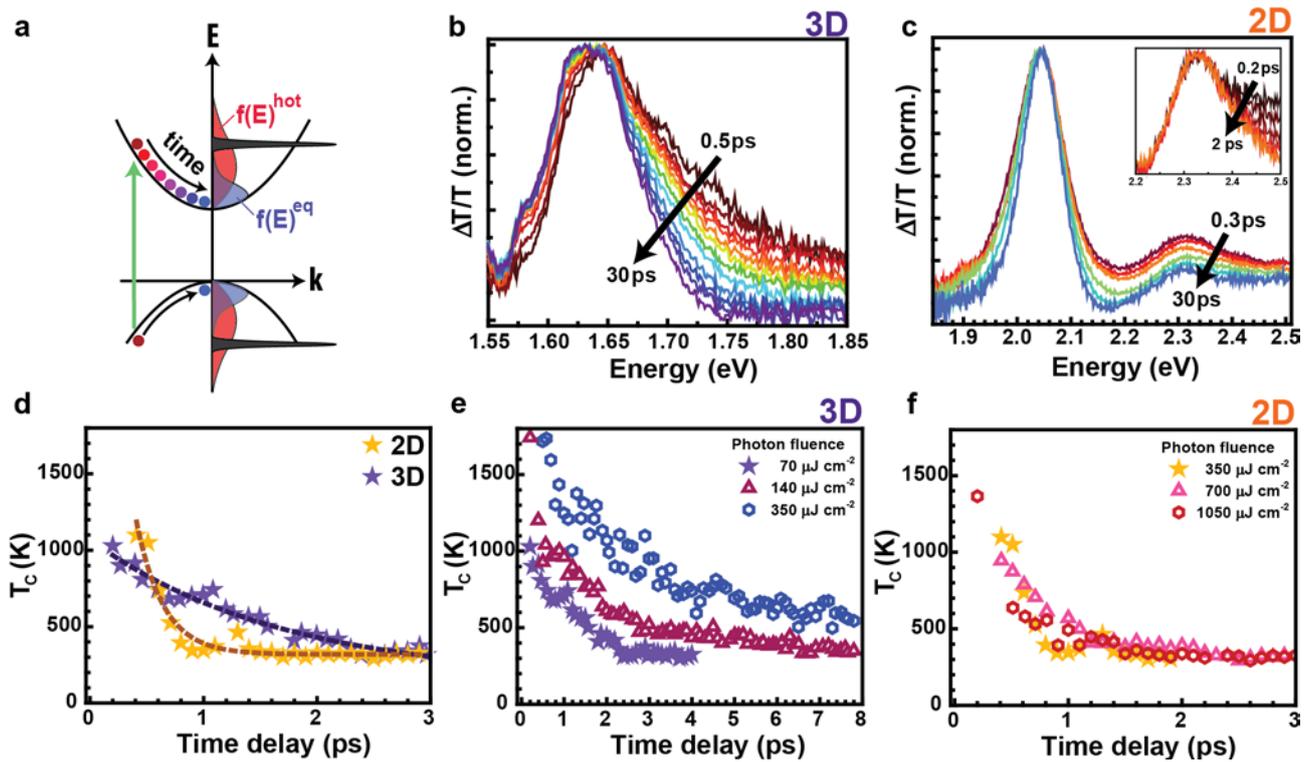

**Figure 2 | Photoexcited charge carrier cooling. a,** Scheme of the thermalisation and relaxation of photo-excited electrons and holes in the one-particle picture. The initially independent δ-like distribution functions are depicted for electrons in the conduction-band and holes in the valence-band. Directly after thermalisation the carriers assume a (hot) thermal distribution $f(E)^{hot}$ (red). $f(E)^{eq}$ (blue) describes the distribution when the charge carriers have cooled down sufficiently to reach equilibrium with the crystal lattice. **b,c,** Normalized ΔT spectra with increasing time delays from 0.3 ps to 30ps for the quasi-3D **(b)** and 2D **(c)** case. In the inset of **(c)** the ΔT-spectra are redrawn normalized to the values at the continuum absorption onset. **d,** Transients of the carrier temperature $T_C$ (cooling curves) as calculated from equ.(3) for the quasi-3D and the 2D sample. The dashed lines represent calculated exponential decays with time constants of 1.7 ps and 240 fs, nicely resembling the experimentally obtained cooling curves. **e,f,** Photon fluence-dependent measurements of the cooling curves for quasi-3D **(e)** and 2D **(f)** NPls.

**Relaxation and cooling**

In order to address the relaxation and cooling dynamics of photoexcited electron-hole pairs, we investigate the DTS signals at time delays between 300 fs and 30 ps. Normalized ΔT spectra are shown in Fig. 2b and 2c for the quasi-3D and 2D cases with photon fluences of 350 µJ/cm² and 700 µJ/cm² per pump pulse, respectively. For all measurements, we assured that the change in absorption upon pumping ($\Delta A/A$) remained below 10% for the two systems in order to be able to compare them and assure that the systems are pumped weakly enough so that the transmission is roughly proportional to the charge carrier density and to remain far away from gain conditions. Apart from the main bleaching signal at the two respective 1s exciton energies (1.62 eV and 2.04 eV) additional transmission changes occur at higher photon energies. For the quasi-3D NPls these extra ΔT-contributions start right on the high energy side of the 1s exciton and exhibit an approximately exponential decay to higher energies (Fig.2b). This is a clear indication that hot electron-hole pair distributions bleach the continuum absorption by phase space filling.[37] For the 2D NPls the corresponding additional ΔT-signal around 2.3 eV is well separated from the 1s exciton signal as can be seen clearly in the inset in Fig. 2c, showing the ΔT-spectra normalized to the values at the continuum absorption onset. These results confirm our interpretation of the absorption spectra within the Wannier-Mott exciton model for the quasi-3D as well as the 2D NPls. For both cases the distribution functions obviously cool down with increasing time delays τ, so we can use the ΔT-spectra above the continuum absorption onset for a given time delay to calculate the respective carrier temperature $T_c$. We assume that the Fermi-Dirac distribution function $f_{FD}(E, T_c)$ can be approximated by a Boltzmann distribution function leading to [38-39]

$$\frac{\Delta T(\tau)}{T} \propto \left(-\frac{E - E_{cont}}{k_B T_c(\tau)}\right)$$

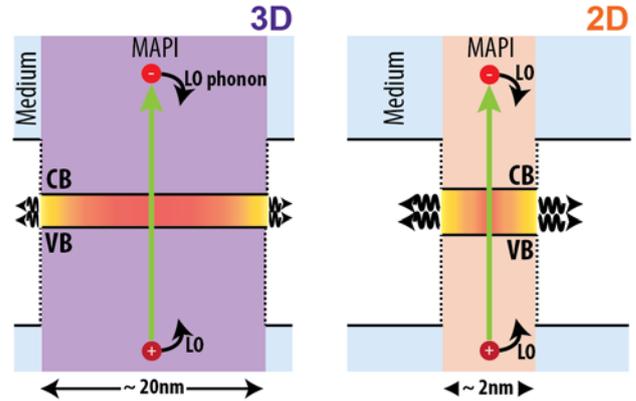

**Figure 3 | Scheme of carrier cooling processes in thick (quasi-3D) and thin (2D) NPls.** After photoexcitation the carriers cool down *via* emission of LO-phonons. As described in the main text, both screening effects and larger surface-to-volume ratios explain why in the 2D case carrier cooling is faster and a dependence on carrier density is not observed.

In this way we obtain transient cooling curves $T_c(\tau)$ for the quasi-3D and the 2D NPls as depicted in Figures 2d-f. To ensure that the carriers have thermalized, we only analyze cooling curves for time delays larger than 200 fs.[32] Excited at the lowest photon fluences possible, the two samples both show initial charge carrier temperatures well above 1000 K as shown in Figure 2d. The charge carriers rapidly cool down in both systems due to emission of optical phonons, which can be modelled by assuming an exponential decay (see dashed lines in Fig.2d). For the quasi-3D system we obtain a time constant of 1.7 ps, while in the 2D system carriers cool down more rapidly, namely with a time constant of 240 fs (see Table S1 for all cooling times). Importantly, these times are significantly faster than the observed Auger recombination (Figure 1c) and also than reported on previously.[40] Consequently, Auger recombination should not affect the charge carrier cooling despite the high laser fluences and can be neglected. This rapid cooling is not followed by a subsequent cooling due to the emission of acoustic phonons, since the thermal energy of the lattice at room temperature ($kT_L = 26\ meV$) exceeds the energy of optical phonons in MAPI ($h\nu_{LO} = 12 - 17\ meV$).[41,42] That the cooling progresses faster in the 2D than in the quasi-3D case seems counterintuitive initially. As the dimensionality is reduced from 3D to 2D, the number of phonon modes and the density of states (DOS) for electrons are reduced. Consequently, one would expect a reduced carrier-phonon scattering rate and therefore longer cooling times for the 2D system. Ultimately, the carrier confinement leads to a so-called phonon-bottleneck and thus slow cooling when the system is completely confined for quantum dots (0D).[43-47] However, for the perovskite NPl system another process leads to enhanced carrier scattering. Instead of being embedded within another semiconductor, the MAPI NPls with a large dielectric permittivity are passivated by organic ligands and immersed in an organic solvent - here toluene - with a small dielectric permittivity, as shown schematically in Figure 3. Consequently, Coulomb interactions with excited charge carriers in the quasi-3D perovskite NPls are screened by the large values of the permittivity, while those in the 2D perovskite NPls will be less screened due to the penetration of electric field lines into the surrounding low dielectric permittivity medium. This not only leads to the extremely large binding energies of excitons but also has a pronounced effect on the scattering of charge carriers among each other and with optical phonons, as governed by the Fröhlich interaction.[33] The observed faster carrier cooling in 2D perovskite NPls is thus a direct consequence of the reduced screening of the Coulomb-mediated carrier-LO-phonon interaction in the low-dielectric surrounding. In turn, the more pronounced screening results in a slower cooling process of carriers in quasi-3D MAPI NPls.

Interestingly, as shown in Figures 2e and 2f, the cooling times for the quasi-3D NPls increase with increasing pump photon fluence, whereas they remain constant for the 2D NPls. Two processes lead to this effect, the first of which we identified as the hot photon effect and has already been observed for bulk perovskites.[38,39,48,49] In this process, the reabsorption of phonons by charge carriers, which depends strongly on the phonon density, slows down the cooling process progressively for increasing photon fluences. Secondly, a higher carrier density increases the dielectric permittivity of the perovskite further, leading to a stronger screening of the Fröhlich interaction and thereby a reduced cooling rate.[19,20,31,35] In the 2D system, exhibiting a large surface-to-volume ratio, emitted optical phonons and subsequently created acoustic phonons have a much larger probability to transfer their energy into the colder medium surrounding the NPls. Thus, heat can dissipate quickly and the probability of hot phonon reabsorption is substantially reduced.[50,51] Furthermore, the increasing photoexcited carrier density can only screen the Fröhlich interaction within the thin 2D NPls, while the field lines of this interaction in the surrounding remain unaffected. Thus the scattering rates stay consistently high even for high excitation densities.

### Resonant *versus* non-resonant excitation

In the experiments presented so far, optical excitation always occurred far above the bandgap, thus creating hot charge carriers, which decay subsequently through phonon emission as depicted in Figure 4a,b. To circumvent the entire cooling process and investigate a (potential) change in the resulting dynamics, we excite the 1s exciton at $E_x = E_C - E_B$ directly (Figure 4d,e). To this end, we pumped the quasi-3D and 2D NPl samples both resonantly with the 1s exciton peaks at 758 nm and 605 nm and non-resonantly at 565 nm and 400 nm, respectively. We then analyze the ΔT-transients for the wavelengths of the probe pulse coinciding with the 1s exciton energy (Figure 4c and 4f). For non-resonant excitation, we see that the ΔT-transient of the quasi-3D sample (Fig.4c, purple data points) shows a fast initial rise followed by a slower increase to the full maximum after 1.8 ps. We ascribe these two distinct dynamical time windows to the carrier thermalisation and the subsequent cooling dynamics of photoexcited carriers as discussed above. Since the 1s exciton binding energy is small for the quasi-3D NPls, carriers reaching the bottom of the continuum transitions are already within the thermal window $kT_L$ of the 1s exciton. In contrast, for non-resonant excitation of the 2D NPls the ΔT-transient detected at the 1s-exciton (Fig.4c, orange data points) shows a significantly slower increase and only one observable time regime. In order to energetically reach the optically accessible $K_{x,y} = 0$ state of the 1s exciton parabola, lying 230 meV below the onset of the continuum transitions, a multitude of LO-phonons (approximately 15 LO-phonons) must be emitted. Due to the small DOS of the excitonic states as compared to the continuum states, this takes time and explains the slow rise of the ΔT-transient for 2D-NPLs. For the quasi-3D NPls the emission of only one LO-phonon is sufficient to overcome this energy difference and only a fraction of photoexcited electron-hole pairs are expected to become converted into 1s excitons ($E_{3D}^B < k_B T_L$). [41,42]

Under resonant excitation (Fig.4d,e), the ΔT-transients of both samples (Fig.4f) increase similarly fast with a time-constant of approximately 200 fs, close to the time-resolution of our experiment. For the quasi-3D NPls the fast onset is followed by a similarly fast decay, which reduces the ΔT-transient down to a value of 70% of the maximum signal in only 170 fs. As explained previously, photoexcited charge carriers initially exhibit δ-function-like nonthermal distributions. As the energy difference between the 1s and continuum is less than the thermal energy and comparable to that of the optical phonons in perovskite, the resonantly created "cold" excitons can be scattered by thermally available phonons into continuum states where the DOS is high. Scattered away from the light line, they no longer bleach the absorption, leading to the rapid drop in the ΔT/T-signal. This dissociation of the exciton has also been observed in four-wave mixing spectroscopy, which we used previously to deduce the homogeneous linewidth of perovskite NCs from the dephasing time $T_2$ of the 1s exciton transition.[24] The subsequent decay, which is significantly slower and constant for all observed times, is due to the recombination of excitons. In contrast, it seems as though the 2D NPls do not exhibit such a behavior. However, if we look at longer timescales of up to 80 ps, as depicted in the inset in Figure 4f and in Figure S4, the decay of the ΔT-transient strongly resembles that of the quasi-3D sample, albeit on a much longer timescale. In fact, the signal decays to approximately 77% of the maximum signal within 10 ps. In principle, the process is the same, as resonantly excited $K_{x,y} = 0$ excitons scatter with phonons (also LO) into $K_{x,y} > 0$ states on the 1s parabola as depicted in Fig.4e. However, with a binding energy larger than 200 meV, excitons cannot be scattered into the continuum and thus thermalisation can only take place with states located on the 1s exciton parabola. As the DOS here is much lower than in the continuum states, the phonon-induced heating of resonantly excited "cold" excitons occurs significantly slower in the 2D than in the quasi 3D-case. Exciton recombination from the radiative window then progresses as discussed above.

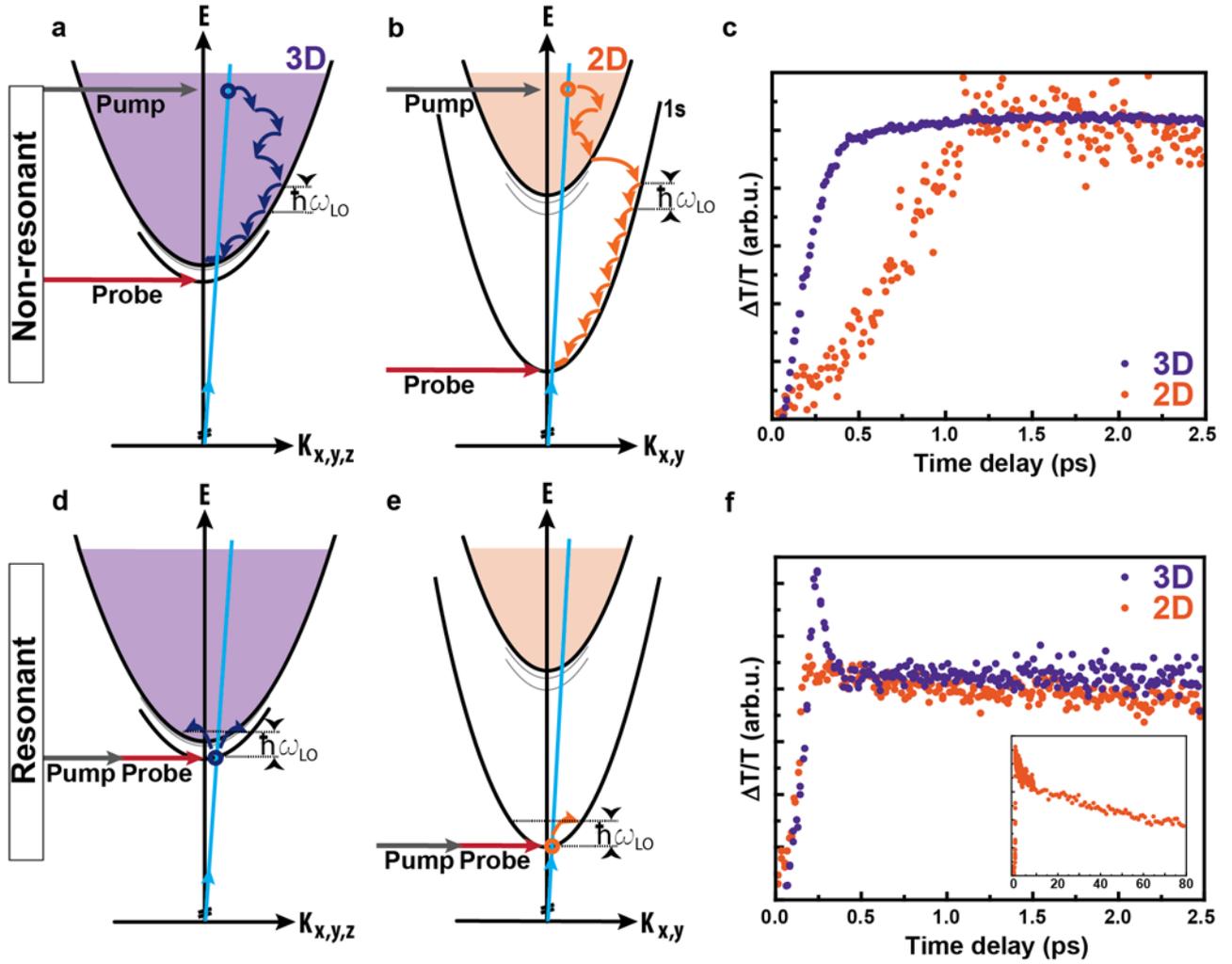

**Figure 4 | Resonant and non-resonant excitation of the 1s exciton transition. a,b** Scheme of charge carrier relaxation after non-resonant (high energy) excitation of the quasi-3D **(a)** and 2D **(b)** NPls. Relaxation occurs predominantly *via* the emission of optical phonons. **c,** Corresponding ΔT-transients for the quasi-3D (purple) and 2D (orange) NPls. **d,e** Scheme of excitonic dynamics after resonant excitation of the 1s exciton transition for quasi-3D **(d)** and 2D **(e)** NPls, respectively. **f,** Corresponding ΔT-transients for the quasi-3D (purple) and 2D (orange) NPls. The inset in **(f)** shows the heating of resonantly excited "cold" 1s excitons.

## **Conclusions**

In summary, we have compared the relaxation and recombination dynamics of photoexcited electron-hole pairs in 2D and 3D MAPI NPls by means of transient DTS. Focusing on the cooling of excited electron-hole pairs, we find striking and initially unexpected differences. For the 2D case relaxation is significantly faster than their 3D counterparts and independent on the excitation fluence. We attribute this to a reduced screening of the Fröhlich interaction in the 2D system and its low dielectric surrounding resulting in enhanced and density-independent phonon emission as well as rapid dissipation of the energy from the phonon system to the surrounding medium. Additionally, we observe a rapid dissociation of "cold" excitons in the 3D system through scattering with optical phonons, an effect which is precluded in the 2D system due to the large exciton binding energy. These results constitute a fundamental insight into how dimensionality affects the relaxation dynamics of unbound and bound electron-hole pairs in halide perovskite NPls. Importantly, these findings will facilitate the development and choice of material parameters for optical devices relying on high speeds and cutoff frequencies, for example photodetectors and all-optical switches. With relaxation times on the order of 0.3 ps, the 2D system could lead to optoelectronic devices with switching times in the THz regime. However, the embedding medium is highly critical to achieving these speeds.

## **Methods**

### **Synthesis of perovskite nanoplatelets (2D and quasi-3D)**

In a typical synthesis, 10 ml of toluene, 0.5 ml of oleylamine and 0.5 ml of oleic acid were added to a mixture of 0.16 mmol MAI and 0.16 mmol PbI2 precursor powders in a 20 ml glass bottle. The reaction mixture was subjected to tip-sonication (SonoPlus HD 3100, Bandelin) at 50% of its maximum power for 30 minutes. During sonication the formation of the perovskite is evident through the change of color of the solution from yellow to black *via* orange and red. The as-obtained solution contains NCs of different sizes and shapes. The as-prepared colloidal solution was centrifuged at a speed of 8000 rpm for 10 min. Then, the supernatant which contained thin plates was separated from the 3D NPls present in the sediment. The sediment was washed with toluene twice and then redispersed in toluene for further studies. The colloidal solution of 3 ML NPls was obtained by adding 200 µl of the initial supernatant to 10 ml of toluene. The resultant colloidal solution appears orange. The morphology of the obtained NPls was characterized with a transmission electron microscope (TEM) operating at an accelerating voltage of 80-100 kV (JEOL JEM-1011). For TEM characterization of 3 ML NPls, 3 ml of the as-obtained orange color solution was centrifuged at a speed of 14000 rpm for 20 min and then the sediment was redispersed in 0.3 mL of hexane.

### **Steady-state optical measurements**

UV-visible absorption spectra of perovskite NPls in solution were recorded using a Varian Cary 5000 UV–vis-IR spectrometer. PL measurements were acquired with a Varian Cary Eclipse fluorescence spectrophotometer (Agilent Technologies).

**Ultrafast transient absorption spectroscopy**

We have employed transient DTS to compare carrier relaxation and recombination dynamics in 2D and quasi-3D halide perovskite NPls. These NPls are dissolved as colloids in toluene and are investigated at room temperature (300 K). Transient differential transmission spectra were taken with a custom built transient absorption spectrometer from Newport Inc. As light source a 1 kHz femtosecond Ti:Sa amplifier system (Libra from Coherent Inc.) was used in combination with an optical parametric amplifier (Opera Solo from Coherent Inc.), providing 100 fs laser pulses with wavelengths over a wide spectral range (290 nm to 10 µm). The diameter of the spot size of the excitation beam was about 350 µm. The same laser system was used to generate a white light probe beam. The samples were measured in ambient conditions in quartz cuvettes with an optical path length of 2 mm and an excitation wavelength of 400 nm (well above the continuum absorption onset). The difference in transmission $\Delta T$ with and without the pump laser pulse present was measured as a function of the time delay $\tau$ between pump and probe pulses. For both samples and for all time delays the differential transmission spectra shown in Fig.2b,c exhibit a maximum bleaching signal located spectrally at the position of the 1s exciton resonance as determined from the linear absorption spectra (Figure 1a and Figure S1 in the Supporting Information). This bleaching reaches a maximum after approximately 1ps and decays subsequently on a much longer time scale.

## References

1. Stranks, S. D.; Eperon, G. E.; Grancini, G.; Menelaou, C.; Alcocer, M. J. P.; Leijtens, T.; Herz, L. M.; Petrozza, A.; Snaith, H. J., Electron-Hole Diffusion Lengths Exceeding 1 Micrometer in an Organometal Trihalide Perovskite Absorber. *Science* **2013,** *342*, 341-344.
2. Zhou, H. P.; Chen, Q.; Li, G.; Luo, S.; Song, T. B.; Duan, H. S.; Hong, Z. R.; You, J. B.; Liu, Y. S.; Yang, Y., Interface Engineering of Highly Efficient Perovskite Solar Cells. *Science* **2014,** *345*, 542-546.
3. De Wolf, S.; Holovsky, J.; Moon, S. J.; Loper, P.; Niesen, B.; Ledinsky, M.; Haug, F. J.; Yum, J. H.; Ballif, C., Organometallic Halide Perovskites: Sharp Optical Absorption Edge and Its Relation to Photovoltaic Performance. *J. Phys. Chem. Lett.* **2014,** *5*, 1035-1039.
4. Saliba, M.; Matsui, T.; Seo, J. Y.; Domanski, K.; Correa-Baena, J. P.; Nazeeruddin, M. K.; Zakeeruddin, S. M.; Tress, W.; Abate, A.; Hagfeldt, A.; Gratzel, M., Cesium-Containing Triple Cation Perovskite Solar Cells: Improved Stability, Reproducibility and High Efficiency. *Energy Environ. Sci.* **2016,** *9*, 1989-1997.
5. Akkerman, Q. A.; D'Innocenzo, V.; Accornero, S.; Scarpellini, A.; Petrozza, A.; Prato, M.; Manna, L., Tuning the Optical Properties of Cesium Lead Halide Perovskite Nanocrystals by Anion Exchange Reactions. *J. Am. Chem. Soc.* **2015,** *137*, 10276-10281.
6. Nedelcu, G.; Protesescu, L.; Yakunin, S.; Bodnarchuk, M. I.; Grotevent, M. J.; Kovalenko, M. V., Fast Anion-Exchange in Highly Luminescent Nanocrystals of Cesium Lead Halide Perovskites ($CsPbX_3$, X = Cl, Br, I). *Nano Lett.* **2015,** *15*, 5635-5640.
7. Xing, G. C.; Mathews, N.; Lim, S. S.; Yantara, N.; Liu, X. F.; Sabba, D.; Gratzel, M.; Mhaisalkar, S.; Sum, T. C., Low-Temperature Solution-Processed Wavelength-Tunable Perovskites for Lasing. *Nat. Mater.* **2014,** *13*, 476-480.
8. Saparov, B.; Mitzi, D. B., Organic-Inorganic Perovskites: Structural Versatility for Functional Materials Design. *Chem. Rev.* **2016,** *116*, 4558-4596.
9. Kovalenko, M. V.; Protesescu, L.; Bodnarchuk, M. I., Properties and Potential Optoelectronic Applications of Lead Halide Perovskite Nanocrystals. *Science* **2017,** *358*, 745-750.
10. Koscher, B. A.; Swabeck, J. K.; Bronstein, N. D.; Alivisatos, A. P., Essentially Trap-Free $CsPbBr_3$ Colloidal Nanocrystals by Postsynthetic Thiocyanate Surface Treatment. *J. Am. Chem. Soc.* **2017,** *139*, 6566-6569.

11. Sichert, J. A.; Tong, Y.; Mutz, N.; Vollmer, M.; Fischer, S.; Milowska, K. Z.; Cortadella, R. G.; Nickel, B.; Cardenas-Daw, C.; Stolarczyk, J. K.; Urban, A. S.; Feldmann, J., Quantum Size Effect in Organometal Halide Perovskite Nanoplatelets. *Nano Lett.* **2015,** *15*, 6521-6527.
12. Hintermayr, V. A.; Richter, A. F.; Ehrat, F.; Döblinger, M.; Vanderlinden, W.; Sichert, J. A.; Tong, Y.; Polavarapu, L.; Feldmann, J.; Urban, A. S., Tuning the Optical Properties of Perovskite Nanoplatelets through Composition and Thickness by Ligand-Assisted Exfoliation. *Adv. Mater.* **2016,** *28*, 9478-9485.
13. Akkerman, Q. A.; Motti, S. G.; Kandada, A. R. S.; Mosconi, E.; D'Innocenzo, V.; Bertoni, G.; Marras, S.; Kamino, B. A.; Miranda, L.; De Angelis, F.; Petrozza, A.; Prato, M.; Manna, L., Solution Synthesis Approach to Colloidal Cesium Lead Halide Perovskite Nanoplatelets with Monolayer-Level Thickness Control. *J. Am. Chem. Soc.* **2016,** *138*, 1010-1016.
14. Liu, J.; Xue, Y.; Wang, Z.; Xu, Z.-Q.; Zheng, C.; Weber, B.; Song, J.; Wang, Y.; Lu, Y.; Zhang, Y.; Bao, Q., Two-Dimensional $CH_3NH_3PbI_3$ Perovskite: Synthesis and Optoelectronic Application. *ACS Nano* **2016,** *10*, 3536-3542.
15. Weidman, M. C.; Goodman, A. J.; Tisdale, W. A., Colloidal Halide Perovskite Nanoplatelets: An Exciting New Class of Semiconductor Nanomaterials. *Chem. Mater.* **2017,** *29*, 5019-5030.
16. Polavarapu, L.; Nickel, B.; Feldmann, J.; Urban, A. S., Advances in Quantum-Confined Perovskite Nanocrystals for Optoelectronics. *Adv. Energy Mater.* **2017,** *7*, 1700267.
17. Wang, Q.; Liu, X.-D.; Qiu, Y.-H.; Chen, K.; Zhou, L.; Wang, Q.-Q., Quantum Confinement Effect and Exciton Binding Energy of Layered Perovskite Nanoplatelets. *AIP Advances* **2018,** *8*, 025108.
18. Yang, D.; Zou, Y.; Li, P.; Liu, Q.; Wu, L.; Hu, H.; Xu, Y.; Sun, B.; Zhang, Q.; Lee, S.-T., Large-Scale Synthesis of Ultrathin Cesium Lead Bromide Perovskite Nanoplates with Precisely Tunable Dimensions and Their Application in Blue Light-Emitting Diodes. *Nano Energy* **2018,** *47*, 235-242.
19. Sohier, T.; Calandra, M.; Mauri, F., Two-Dimensional Frohlich Interaction in Transition-Metal Dichalcogenide Monolayers: Theoretical Modeling and First-Principles Calculations. *Phys. Rev. B* **2017,** *96*, 085415.
20. Shah, J., *Hot Carriers in Semiconductor Nanostructures: Physics and Applications*. Academic Press, Inc.: San Diego, CA, 1992.
21. Ma, J.; Wang, L. W., Nanoscale Charge Localization Induced by Random Orientations of Organic Molecules in Hybrid Perovskite $CH_3NH_3PbI_3$. *Nano Lett.* **2015,** *15*, 248-253.
22. Bohn, B. J.; Tong, Y.; Gramlich, M.; Lai, M. L.; Döblinger, M.; Wang, K.; Hoye, R. L. Z.; Müller-Buschbaum, P.; Stranks, S. D.; Urban, A. S.; Polavarapu, L.; Feldmann, J., Boosting Tunable Blue Luminescence of Halide Perovskite Nanoplatelets through Postsynthetic Surface Trap Repair. *Nano Lett.* **2018,** *18*, 5231-5238.
23. Herz, L. M., Charge-Carrier Dynamics in Organic-Inorganic Metal Halide Perovskites. *Annu. Rev. Phys. Chem.* **2016,** *67*, 65-89.
24. Bohn, B. J.; Simon, T.; Gramlich, M.; Richter, A. F.; Polavarapu, L.; Urban, A. S.; Feldmann, J., Dephasing and Quantum Beating of Excitons in Methylammonium Lead Iodide Perovskite Nanoplatelets. *ACS Photonics* **2018,** *5*, 648-654.
25. Miyata, A.; Mitioglu, A.; Plochocka, P.; Portugall, O.; Wang, J. T.-W.; Stranks, S. D.; Snaith, H. J.; Nicholas, R. J., Direct Measurement of the Exciton Binding Energy and Effective Masses for Charge Carriers in Organic–Inorganic Tri-Halide Perovskites. *Nat. Phys.* **2015,** *11*, 582-587.
26. Ishihara, T.; Takahashi, J.; Goto, T., Optical Properties Due to Electronic Transitions in Two-Dimensional Semiconductors $(C_nH_{2n+1}NH_3)_2PbI_4$. *Phys. Rev. B* **1990,** *42*, 11099-11107.
27. Hong, X.; Ishihara, T.; Nurmikko, A. V., Dielectric Confinement Effect on Excitons in $PbI_4$-Based Layered Semiconductors. *Phys. Rev. B* **1992,** *45*, 6961-6964.
28. Hirasawa, M.; Ishihara, T.; Goto, T.; Sasaki, S.; Uchida, K.; Miura, N., Magnetoreflection of the Lowest Exciton in a Layered Perovskite-Type Compound $(C_{10}H_{21}NH_3)_2PbI_4$. *Solid State Commun.* **1993,** *86*, 479-483.
29. Feldmann, J.; Peter, G.; Gobel, E. O.; Dawson, P.; Moore, K.; Foxon, C.; Elliott, R. J., Linewidth Dependence of Radiative Exciton Lifetimes in Quantum-Wells. *Phys. Rev. Lett.* **1987,** *59*, 2337-2340.

30. Tessier, M. D.; Javaux, C.; Maksimovic, I.; Loriette, V.; Dubertret, B., Spectroscopy of Single CdSe Nanoplatelets. *ACS Nano* **2012,** *6*, 6751-6758.
31. Shah, J., *Ultrafast Spectroscopy of Semiconductors and Semiconductor Nanostructures*. Springer: Berlin ; New York, 1996; p xv, 372 p.
32. Richter, J. M.; Branchi, F.; Camargo, F. V. D.; Zhao, B. D.; Friend, R. H.; Cerullo, G.; Deschler, F., Ultrafast Carrier Thermalization in Lead Iodide Perovskite Probed with Two-Dimensional Electronic Spectroscopy. *Nat. Commun.* **2017,** *8*, 376.
33. Potz, W., Hot-Phonon Effects in Bulk GaAs. *Phys. Rev. B* **1987,** *36*, 5016-5019.
34. Brinkman, W. F.; Rice, T. M., Electron-Hole Liquids in Semiconductors. *Phys. Rev. B* **1973,** *7*, 1508-1523.
35. Skolnick, M. S.; Nash, K. J.; Tapster, P. R.; Mowbray, D. J.; Bass, S. J.; Pitt, A. D., Free-Carrier Screening of the Interaction between Excitons and Longitudinal-Optical Phonons in $In_xGa_{1-x}As$-InP Quantum-Wells. *Phys. Rev. B* **1987,** *35*, 5925-5928.
36. Kaasbjerg, K.; Bhargavi, K. S.; Kubakaddi, S. S., Hot-Electron Cooling by Acoustic and Optical Phonons in Monolayers of $MoS_2$ and Other Transition-Metal Dichalcogenides. *Phys. Rev. B* **2014,** *90*.
37. Manser, J. S.; Kamat, P. V., Band Filling with Free Charge Carriers in Organonietal Halide Perovskites. *Nat. Photonics* **2014,** *8*, 737-743.
38. Price, M. B.; Butkus, J.; Jellicoe, T. C.; Sadhanala, A.; Briane, A.; Halpert, J. E.; Broch, K.; Hodgkiss, J. M.; Friend, R. H.; Deschler, F., Hot-Carrier Cooling and Photoinduced Refractive Index Changes in Organic-Inorganic Lead Halide Perovskites. *Nat. Commun.* **2015,** *6*, 8420.
39. Yang, Y.; Ostrowski, D. P.; France, R. M.; Zhu, K.; van de Lagemaat, J.; Luther, J. M.; Beard, M. C., Observation of a Hot-Phonon Bottleneck in Lead-Iodide Perovskites. *Nat. Photonics* **2016,** *10*, 53-59.
40. Eperon, G. E.; Jedlicka, E.; Ginger, D. S., Biexciton Auger Recombination Differs in Hybrid and Inorganic Halide Perovskite Quantum Dots. *J. Phys. Chem. Lett.* **2018,** *9*, 104-109.
41. Wright, A. D.; Verdi, C.; Milot, R. L.; Eperon, G. E.; Perez-Osorio, M. A.; Snaith, H. J.; Giustino, F.; Johnston, M. B.; Herz, L. M., Electron-Phonon Coupling in Hybrid Lead Halide Perovskites. *Nat. Commun.* **2016,** *7*, 11755.
42. Sendner, M.; Nayak, P. K.; Egger, D. A.; Beck, S.; Müller, C.; Epding, B.; Kowalsky, W.; Kronik, L.; Snaith, H. J.; Pucci, A.; Lovrinčić, R., Optical Phonons in Methylammonium Lead Halide Perovskites and Implications for Charge Transport. *Mater. Horizons* **2016,** *3*, 613-620.
43. Heitz, R.; Kalburge, A.; Xie, Q.; Grundmann, M.; Chen, P.; Hoffmann, A.; Madhukar, A.; Bimberg, D., Excited States and Energy Relaxation in Stacked InAs/GaAs Quantum Dots. *Phys. Rev. B* **1998,** *57*, 9050-9060.
44. Guyot-Sionnest, P.; Shim, M.; Matranga, C.; Hines, M., Intraband Relaxation in CdSe Quantum Dots. *Phys. Rev. B* **1999,** *60*, R2181-R2184.
45. Pandey, A.; Guyot-Sionnest, P., Slow Electron Cooling in Colloidal Quantum Dots. *Science* **2008,** *322*, 929-932.
46. Feldmann, J.; Cundiff, S. T.; Arzberger, M.; Bohm, G.; Abstreiter, G., Carrier Capture into InAs/GaAs Quantum Dots *via* Multiple Optical Phonon Emission. *J. Appl. Phys.* **2001,** *89*, 1180-1183.
47. Li, M. J.; Bhaumik, S.; Goh, T. W.; Kumar, M. S.; Yantara, N.; Gratzel, M.; Mhaisalkar, S.; Mathews, N.; Sum, T. C., Slow Cooling and Highly Efficient Extraction of Hot Carriers in Colloidal Perovskite Nanocrystals. *Nat. Commun.* **2017,** *8*, 14350.
48. Deschler, F.; Price, M.; Pathak, S.; Klintberg, L. E.; Jarausch, D. D.; Higler, R.; Huttner, S.; Leijtens, T.; Stranks, S. D.; Snaith, H. J.; Atature, M.; Phillips, R. T.; Friend, R. H., High Photoluminescence Efficiency and Optically Pumped Lasing in Solution-Processed Mixed Halide Perovskite Semiconductors. *J. Phys. Chem. Lett.* **2014,** *5*, 1421-1426.
49. Fu, J. H.; Xu, Q.; Han, G. F.; Wu, B.; Huan, C. H. A.; Leek, M. L.; Sum, T. C., Hot Carrier Cooling Mechanisms in Halide Perovskites. *Nat. Commun.* **2017,** *8*, 1300.
50. Bao, H.; Habenicht, B. F.; Prezhdo, O. V.; Ruan, X. L., Temperature Dependence of Hot-Carrier Relaxation in Pbse Nanocrystals: An *Ab Initio* Study. *Phys. Rev. B* **2009,** *79*, 235306.
51. Cooney, R. R.; Sewall, S. L.; Dias, E. A.; Sagar, D. M.; Anderson, K. E. H.; Kambhampati, P., Unified Picture of Electron and Hole Relaxation Pathways in Semiconductor Quantum Dots. *Phys. Rev. B* **2007,** *75*, 245311.

## Associated Content

**Supporting Information**

The Supporting Information is available free of charge on the ACS Publications website at DOI:xx.xxxx/acsnano.xxxxxxx. Detailed information on the applied Elliot model and fitting curves of the model applied to linear absorption spectra, TEM images of the perovskite NPls, Laser fluences and corresponding charge carrier cooling times, Differential transmission curves at maximum bleach signal and resonant excitation of the 1s state.

**Financial Information**

The authors declare no competing financial interest.

## Author Information

**Corresponding Authors**

* Email: urban@lmu.de

* Email: feldmann@lmu.de

**ORCID**

Lakshminarayana Polavarapu: 0000-0002-9040-5719

Alexander S. Urban: 0000-0001-6168-2509

## Acknowledgements

This work was supported by the Bavarian State Ministry of Science, Research, and Arts through the grant "Solar Technologies go Hybrid (SolTech)", by the European Research Council Horizon 2020 Marie Skłodowska-Curie Grant Agreement COMPASS (691185) and the ERC Grant Agreement PINNACLE (759744), and by LMU Munich's Institutional Strategy LMUexcellent within the framework of the German Excellence Initiative. The authors would like to thank Thomas Simon for helpful discussions.